\documentclass[aps,prf,onecolumn,superscriptaddress,notitlepage]{revtex4-1}
\usepackage{graphics}
\usepackage{graphicx}
\usepackage{color}
\usepackage{epstopdf}
\usepackage{amsmath}
\usepackage{amssymb}
\usepackage{xcolor}
\usepackage{float} 
\usepackage[caption = false]{subfig}
\usepackage{appendix}
\begin{document}
\title{Capillary Levelling of Immiscible Bilayer Films}
\author{Vincent Bertin}
\thanks{These authors contributed equally to this work.}
\affiliation{Univ. Bordeaux, CNRS, LOMA, UMR 5798, 33405 Talence, France.}
\affiliation{UMR CNRS Gulliver 7083, ESPCI Paris, PSL Research University, 75005 Paris, France.}
\author{Carmen L. Lee}
\thanks{These authors contributed equally to this work.}
\affiliation{Department of Physics and Astronomy, McMaster University, 1280 Main Street West, Hamilton, Ontario, L8S 4M1, Canada.} 
\author{Thomas Salez}
\affiliation{Univ. Bordeaux, CNRS, LOMA, UMR 5798, 33405 Talence, France.}
\affiliation{Global Station for Soft Matter, Global Institution for Collaborative Research and Education, Hokkaido University, Sapporo, Japan.}
\author{Elie Rapha\"el}
\affiliation{UMR CNRS Gulliver 7083, ESPCI Paris, PSL Research University, 75005 Paris, France.}
\author{Kari Dalnoki-Veress}
\email{dalnoki@mcmaster.ca}
\affiliation{UMR CNRS Gulliver 7083, ESPCI Paris, PSL Research University, 75005 Paris, France.}
\affiliation{Department of Physics and Astronomy, McMaster University, 1280 Main Street West, Hamilton, Ontario, L8S 4M1, Canada.} 
\begin{abstract}
Flow in thin films is highly dependent on the boundary conditions. Here, we study the capillary levelling of thin bilayer films composed of two immiscible liquids. Specifically, a stepped polymer layer is placed atop another, flat polymer layer. The Laplace pressure gradient resulting from the curvature of the step induces flow in both layers, which dissipates the excess capillary energy stored in the stepped interface. The effect of different viscosity ratios between the bottom and top layers is investigated. We invoke a long-wave expansion of low-Reynolds-number hydrodynamics to model the energy dissipation due to the coupled viscous flows in the two layers. Good agreement is found between the experiments and the model. Analysis of the latter further reveals an interesting double crossover in time, from Poiseuille flow, to plug flow, and finally to Couette flow. The crossover time scales depend on the viscosity ratio between the two liquids, allowing for the dissipation mechanisms to be selected and finely tuned by varying this ratio.
\end{abstract}
\maketitle

\date{}
\maketitle
\section{Introduction}
Flow in a thin film is affected by the boundary conditions of the film, especially when the thickness of the film approaches that of the boundary layer~\cite{OronRev}. As an example, the presence of slippage at a solid-liquid interface affects flows in thin films as observed in the dewetting dynamics of thin polymer films~\cite{kargupta2004instability, fetzer2005new, munch2005lubrication, baumchen2009slip}. The dynamics is more complex in bilayer or stratified films, because the flow depends on the relative viscosities and interfacial energies of the two layers in addition to the interfacial boundary conditions~\cite{wyart1993liquid, pototsky2004alternative, MerabiaDewetSim, jachalski2014}. Liquid-liquid interfaces, in particular those between two polymers, often exhibit apparent slip~\cite{deGennes1989,deGennesBrochard1990,deGennesBrochard1993}, and have been studied with molecular dynamics simulations~\cite{Koplik, razavi2014molecular, poesio2017slip} and experiments~\cite{lee2009polymer, XuSlip}. Such an effective reduction of friction has important practical implications, \textit{e.g.} smart liquid-impregnated surfaces~\cite{Howell2015,Keiser2017}. The stability and dewetting of thin multilayer polymer films is also a subject of interest for physicists~\cite{lambooy1996dewetting, segalman1999dynamics, lal2017dewetting, peschka2018impact}, and finds applications in industry \textit{e.g.} materials manufactured from coextrusion processes~\cite{ZhaoExtrude, ponting2010polymer,Bironeau2017,chebil2018influence}.
 
Capillary-driven levelling occurs when an excess of interfacial area is relaxed by smoothing topographical perturbations, such as a thin film with some surface feature: a bump, a valley, a hole, etc. Typically, the levelling is driven by the surface tension $\gamma$ of the liquid-vapour interface. The curvature of the free interface results in a Laplace pressure, and a gradient in the curvature induces flow, thereby reducing the surface energy of the system. The flow is mediated by the viscosity $\eta$ of the liquid. Capillary-driven levelling is a useful tool for studying fluid flow in nanofilms and can be used to investigate the boundary conditions~\cite{deGenneBrochardWyartQuereCapillarity}. With well-known initial conditions, capillary-driven levelling has been used to study various interfacial polymeric properties, such as glass transition anomalies, confinement effects, and nanorheology in thin polymers films~\cite{buck2004decay, FakhraaiTg, Yang2010,  Teisseire2011, RogninWave, Chai}. Previous work on nanorheology in thin films has shown that, in addition to the importance of surface tension and viscosity, the flow is sensitive to the boundary conditions~\cite{XuUnder, munch2005lubrication, jachalski2014}. The capillary-levelling technique was applied to a variety of geometries and configurations, which range from imprinted nano-patterns~\cite{stillwagon1988,buck2004decay,RogninWave,Teisseire2011}, to steps~\cite{McGrawViscosity}, trenches~\cite{BaumchenTrench}, holes~\cite{BackholmCylinder,bertin2019symmetrization}, and inhomogeneous mixtures~\cite{McGraw2013}.

\begin{figure}[t]
\centering
\subfloat{\includegraphics[width = \textwidth]{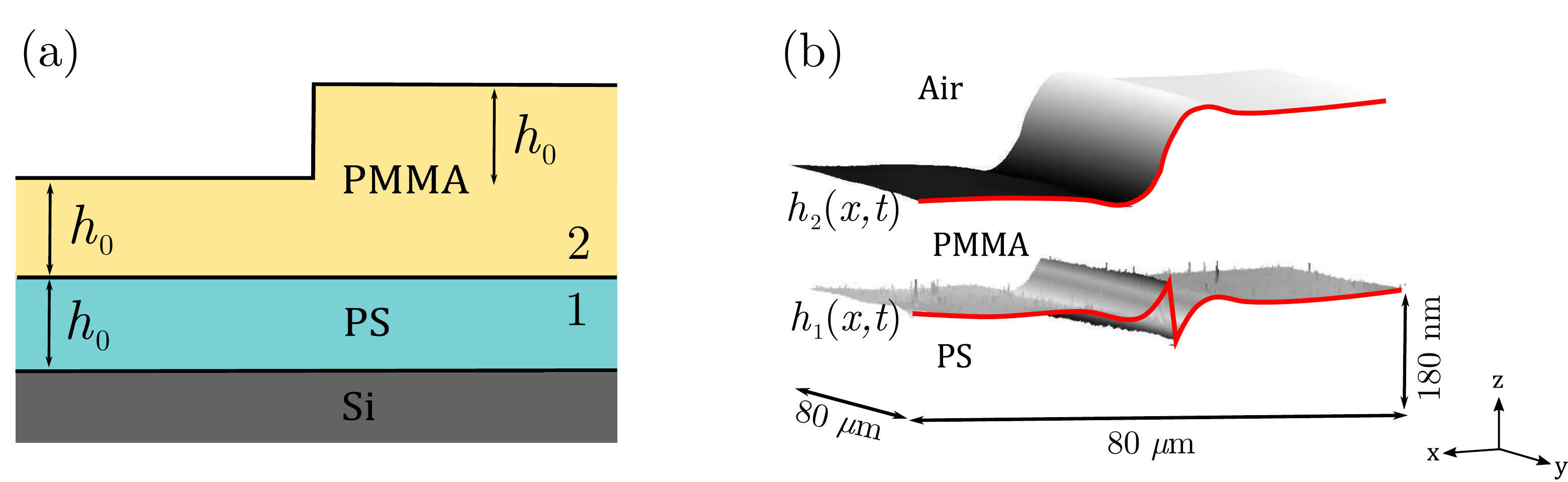}}
\caption{(a) Schematic of the as-prepared sample, with the indicated materials. The upper portion of the stepped film, the lower portion of the stepped film, and the bottom film all have the same thickness $h_{0}$. (b) Atomic force microscopy profiles of the liquid-air interface and the liquid-liquid interface. The two scans are shifted in the vertical direction to reconstruct the actual profile of the  sample.}
\label{fig:initcond}
\end{figure}

In the present work, we focus on the influence of a deformable liquid-liquid interface between two immiscible polymers by studying the capillary-driven levelling of a stepped bilayer film. The latter is depicted schematically in Fig.~\ref{fig:initcond}(a). A stepped polymer film is placed on a flat film of another, immiscible polymer supported on a rigid substrate. The initial surface perturbation can be described as a Heaviside function, where the vertical height profile varies abruptly from one thickness to another as the horizontal $x$-direction is varied. The system is invariant in the other horizontal direction. During the subsequent evolution, the height profile $h(x,t)$ can be described as a function of both the horizontal position $x$ and time $t$. Furthermore, the dynamics is expected to depend on the relative viscosities of the bilayer. Indeed, one can expect that if the viscosity of the bottom flat film is much higher than that of the top stepped film, then the former is much like a rigid substrate: the top film can flow like a liquid film on a solid substrate. In contrast, if the bottom film has a relatively negligible viscosity, then the top film can flow with little hindrance at the bottom, akin to a freestanding liquid film. For these reasons, it is of value to consider the two extreme cases of a single film on a solid substrate and a freestanding film.

In the case of a thin liquid film on a solid substrate~\cite{McGrawViscosity, McGrawEnergy, SalezAnalytic}, there is typically a no-slip boundary condition at the solid-liquid interface and a no-shear-stress boundary condition at the liquid-air interface. Using the lubrication approximation for Stokes flow, the interface profile follows the thin film equation~\cite{OronRev} with a parabolic Poiseuille velocity profile. In earlier works on stepped films, it was found that the thin film equation admits a self-similar solution in the rescaled variable $x/t^{1/4}$~\cite{McGrawEnergy, SalezAnalytic}. 

In contrast, for a freestanding film, there are no-shear-stress boundary conditions at each of the two interfaces. As a consequence, the excess surface energy of a symmetric topographical perturbation must be dissipated through elongational flow, instead of shear flow, as was found in soap films~\cite{Acheson}. Within a long-wave approximation, the flow profile is consistent with plug flow. The interface profile $h(x,t)$ follows a system of coupled partial differential equations~\cite{erneux1993nonlinear} which admits a self-similar solution in the rescaled variable $x/t^{1/2}$~\cite{IltonFreestanding}. We note that freestanding films are described by the same equations as that for supported films on slippery substrates with an \emph{infinite slip length}, since the absence of friction at the solid-liquid interface implies the absence of any shear stress at that interface~\cite{munch2005lubrication}. 

For a thin liquid film placed onto another, immiscible thin liquid film, the flow profile  depends on the viscosity ratio between the two films, as well as on the ratio between the two relevant interfacial tensions. While the levelling of a liquid film atop a more viscous liquid is expected to be similar to that of a liquid film atop a solid substrate, the opposite case of liquid film atop a lower-viscosity liquid is non-trivial and is the primary focus of the present article. Here, we use experiments and low-Reynolds-number hydrodynamics in order to investigate the flow in such a geometry. Based on previous works on supported and freestanding films, we expect the widths of the interfacial perturbations to follow some combinations of the limiting $\sim t^{1/4}$ and $\sim t^{1/2}$ relaxation laws. We demonstrate that the main viscous dissipative mechanism crossovers in time from being Poiseuille-like, to elongational, and then to Couette-like, and that this double crossover is tuneable with the viscosity ratio.

\section{Methods}
\subsection{Experiments}
The sample preparation and experimental protocol follow modified versions of the ones described in previous works~\cite{McGrawViscosity,peschka2018impact}. Figure~\ref{fig:initcond}(a) shows a schematic of the sample, with a stepped film of poly(methyl methacrylate) (PMMA) atop a polystyrene (PS) film supported on a silicon (Si) substrate. PS and PMMA are an immiscible pair~\cite{TanakaMacromol1996}. The thin polymer films are prepared by spin coating PS or PMMA from solutions in toluene (Fisher Chemical, Optima), onto 1~cm $\times$ 1 cm silicon (Si) wafers (University Wafer) and freshly cleaved mica sheets (Ted Pella, Inc.), respectively. The PMMA molecular weight is $M_{\textrm{w}} =$~56~kg/mol (Polymer Source, Inc., polydispersity index $\leq$1.08). The PS molecular weights are $M_{\textrm{w}} =$~53.3, 183, and 758.9 kg/mol (Polymer Source, Inc. and Scientific Polymer Products, Inc., polydispersity index $\leq$1.06). After spin coating, all films are annealed at 150$^{\circ}$C, \textit{i.e.} above the glass transition temperatures of both polymers, for 10 minutes to remove any residual solvent and to relax the polymer chains. The films of PMMA are then floated from the mica surface onto a bath of ultra-pure water (18.2 M$\Omega\cdot$cm). A floating film of PMMA is transferred off the water surface onto the PS-coated silicon substrate to create a flat bilayer film supported on Si. The molecular weight of the PMMA is low enough such that the polymer chains are not highly entangled: a thin film of this polymer easily fractures upon perturbation on the surface of water, which results in portions with straight edges~\cite{BaumchenTrench}. A second film of PMMA is fractured and transferred onto the flat bilayer, the sharp, fractured edge thus creating the step (see Fig.~\ref{fig:initcond}(a)). For each sample studied here, the upper portion of the PMMA stepped film, the lower portion of the PMMA stepped film, and the bottom PS film all have the same thickness, $h_{0}$, within 10\% of each other as confirmed with ellipsometry (Accurion, EP3). The thicknesses used in this work are $h_{0}=100$~nm, 180~nm, and 240~nm.

To examine the evolution of the step with time, the samples are annealed above the glass transition temperature of both polymers ($\sim 100^\circ$C), with a temperature controlled stage (Linkam). The samples are held at the elevated temperature for a given period of time, during which flow and levelling occur, before being quenched back into the glassy state at room temperature. Surface profiles of the liquid-air interface are obtained with atomic force microscopy (AFM, Bruker). For some experiments, the liquid-liquid interface is exposed by dissolving of the top PMMA layer with a selective solvent ($\sim$ 67\% acetic acid and $\sim$ 33\% ultra-pure water). This procedure allows for AFM profiles of the liquid-liquid interface to be measured. Figure~\ref{fig:initcond}(b) shows typical AFM profiles of the liquid-air and liquid-liquid interfaces \emph{taken at the same location on the sample}. The actual profile of the whole sample is reconstructed by vertically shifting the AFM profiles according to the original layer thickness.

The dynamics of capillary-driven levelling depends on two dimensionless numbers: the viscosity ratio $\mu = \eta_1/\eta_2$ between the bottom (PS, 1) and top (PMMA, 2) layers, and the interfacial tension ratio $\Gamma = \gamma_1/\gamma_2$ between the liquid-liquid and liquid-air interfaces. The viscosity ratio is varied by changing the molecular weight of the bottom layer, as well as by changing the annealing temperature; this results in the viscosity ratio varying over 6 orders of magnitude, from $\mu \approx 5.3 \times 10^{-5}$ to $\mu \approx 1.4 \times 10^{1}$. The individual viscosities were measured independently through the capillary levelling of simple stepped films of each single polymer, using the method described previously~\cite{McGrawViscosity}(see Tab.~\ref{table}). Variations of the interfacial tension ratio over the experimental temperature range are negligible, so that the ratio is taken to be $\Gamma = 0.053$~\cite{WuGamma}; thus the liquid-air surface tension largely dominates that of the liquid-liquid interface.
\begin{table}[hb!]
\begin{tabular}{|c|c|c|c|}
\hline
$M_{\textrm{w}}({\mathrm{PS}})$ & $T$ & $\mu$ & $\mu$  \\ 
(kg/mol) & ($^\circ$C) & (best fit) & (independent) \\ \hline
53.3 & 150 & 1.1 $\times 10^{-4}$ & 5.3 $\times 10^{-5}$ \\ \hline
53.3 & 165 & 1.5 $\times 10^{-3}$ & 3.0 $\times 10^{-4}$ \\ \hline
53.3 & 180 & 7.1 $\times 10^{-3}$ & 7.1 $\times 10^{-4}$ \\ \hline
183 &150 & 8.5 $\times 10^{-3}$ & 4.2 $\times 10^{-3}$ \\ \hline
183 & 165 & 7.4 $\times 10^{-3}$ & 2.0 $\times 10^{-2}$ \\ \hline
183 &180 & 4.6 $\times 10^{-2}$ & 4.6 $\times 10^{-2}$ \\ \hline
758.9 & 150 & 1.7 $\times 10^{-1}$ & 6.9 $\times 10^{-1}$ \\ \hline
758.9 & 165 & 1.2 $\times 10^{0}$ & 5.8 $\times 10^{0}$ \\ \hline
758.9 & 180 & 1.5 $\times 10^{0}$ & 1.4 $\times 10^{1}$ \\ \hline
\end{tabular}
\caption{The viscosity ratios $\mu$, between the bottom (PS) and top (PMMA) layers, for various PS molecular weights $M_{\textrm{w}}({\mathrm{PS}})$, and annealing temperatures $T$. The viscosity ratios are obtained from: i) a best fit of the theory to the experimental excess capillary energy (``best fit"); and ii) the capillary levelling of simple stepped films of each single polymer (``independent").}
\label{table}
\end{table}

\subsection{Theory}
\label{subsec:theory}
The system is modelled as two thin liquid layers atop each other, the ensemble being placed on a rigid substrate, as sketched in Fig.~\ref{fig:initcond}(a), and Cartesian coordinates ($x$, $y$, $z$) are used, as shown in Fig.~\ref{fig:initcond}(b). The system is assumed to be infinite in both the $x$-direction and $y$-direction, and invariant by translation in the latter. The typical length scales of the experiment are well below the capillary length, thus gravitational effects can be neglected. IIn thin, highly viscous polymer films, with Reynolds and Mach numbers are much smaller than 1, relaxation is driven by capillarity, and inertial and compressibility effects can be neglected.  Furthermore, the polymer melts may be treated as Newtonian liquids~\cite{McGrawEnergy}, since the typical viscoelastic times, under the present experimental conditions, are on the order of a few seconds~\cite{hirai2003mechanical} whereas the levelling time scales are much larger (minutes to hours). Finally, the film thicknesses are chosen to be large enough such that disjoining forces are weak in comparison with the Laplace pressure.

The velocity fields and excess pressure fields with respect to the atmospheric pressure are denoted as $\vec{u_i} = (u_i, 0, w_i)$ and $p_i$, respectively. In the small-slope limit, the tangential stress balance at the liquid-liquid interface reads $\eta_1 \frac{\partial u_1}{\partial z} = \eta_2 \frac{\partial u_2}{\partial z}$~\cite{jachalski2014}. In the regime where $\mu \ll 1$, this relation further leads to  $\frac{\partial u_2}{\partial z} = 0$ to leading order. Together with a no-shear-stress boundary condition at the liquid-air interface, these are consistent with plug flow in the top layer: much like in the freestanding case discussed above. Within the lubrication approximation, the bottom layer is expected to display a horizontal Poiseuille-like flow. Furthermore, we assume continuity of the velocity field across the liquid-liquid interface, \textit{i.e.} we impose a no-slip boundary condition. As a result, one expects an additional linear term in $z$ (like Couette flow for a simple shear geometry) in the horizontal velocity field of the bottom layer. 

Within this framework, and invoking the lubrication-like scale separation, the heights $h_i(x,t)$ of both interfaces (see Fig.~\ref{fig:initcond}(b)) follow a set of non-linear partial differential equations (see Appendix~\ref{AppendixAsymptotic} for more details). We refer to this first model as the \emph{asymptotic model}. We note that a similar derivation was made for the non-Newtonian case for the upper liquid using the Jeffreys model~\citep{jachalski2015thin}. The governing equations are:
\begin{subequations}
\begin{equation}
\label{GovEq1}
\partial_t (h_2 - h_1) = -\bigg[ (h_2 - h_1)u_2 \bigg]'\ ,
\end{equation}
\begin{equation}
\label{GovEq2}
\partial_t h_1 = -\bigg[(\gamma_2 h_2'''+\gamma_1 h_1''')\frac{h_1^3}{12\eta_1} +\frac{h_1 u_2}{2}   \bigg]' = -\bigg[-p_1'\frac{h_1^3}{12\eta_1} +\frac{h_1 u_2}{2}   \bigg]' \ ,
\end{equation}
\begin{equation}
\label{GovEq3}
\gamma_2 h_2''' (h_2 - h_1) + (\gamma_2 h_2''' + \gamma_1 h_1''') \frac{h_1}{2}  + 4 \eta_2 \bigg[u_2' (h_2-h_1)\bigg]' - \eta_1 \frac{u_2}{h_1} = 0 \ ,
\end{equation}
\end{subequations}
where the prime indicates a derivative with respect to $x$. The excess pressure field $p_1 = -\gamma_1 h_1''(x) - \gamma_2 h_2''(x)$ in the bottom film corresponds to the sum of the two interfacial Laplace pressure jumps in the small-slope limit. Notably, Eq.~(\ref{GovEq3}) has the same form as the tangential stress balance for a single liquid film on a solid substrate with a large slip length~\cite{munch2005lubrication}. The associated apparent slip length in our configuration is $b \sim  h_1 / \mu$~\cite{jachalski2015thin}, which is large if $\mu \ll 1$, \textit{i.e.} if the bottom layer is much less viscous than the top one. We note that a similar analogy with flow over a slippery substrate has been proposed to describe the flow of nanobubbles on hydrophobic surfaces~\cite{LaugaBrenner2004}.

The heights of the two interfaces can be further expressed as perturbations from the equilibrium configuration: $h_1(x,t) = \bar{h}_1 + \delta h_1(x,t)$ and $h_2(x,t) = \bar{h}_2 + \delta h_2 (x,t)$, where $\bar{h}_i$ denote the mean heights of the two interfaces: $\bar{h}_1= h_0$ and $\bar{h}_2= 5h_0/2$ in our specific geometry. We then assume that $\delta h_i \ll \bar{h}_i$, and keep only the leading-order linear terms. We stress that this condition is not strictly valid at the liquid-air interface, but: i) the linearization allows one to obtain an analytical solution; and ii) in both limiting cases of freestanding and supported films, the linearization does preserve the self-similar structure of the non-linear problem~\cite{SalezAnalytic,Salez2012,IltonFreestanding}. Therefore, our approach is still expected to provide some relevant features for the experimental system.

Using the  Fourier transform $\tilde{f}(k)$ of a function $f(x)$, defined as $\tilde{f}(k) = \frac{1}{\sqrt{2 \pi}} \int \textrm{d}x f(x) \exp(i k x)$, we find from the linearization of the governing equations above, that:
\begin{equation}
\label{LinearizedGoverningEquation}
\frac{\partial \tilde{\delta h_i}}{\partial t} = s_{\textrm{i},\textrm{j}}(k) \tilde{\delta h_j}\ ,
\end{equation}
with $s_{\textrm{i},\textrm{j}}$ representing the elements of the decay-rate matrix $\mathbf{s}$ associated with the mode $k$ (see Appendix~\ref{AppendixAsymptotic}). The Einstein summation convention is used in Eq.~\eqref{LinearizedGoverningEquation}. The general solution to this set of equations is: 
\begin{subequations}
\label{eq:solution}
\begin{equation}
\tilde{\delta h_1} = \alpha \exp{(\lambda_1 t)} + \beta \exp{(\lambda_2 t)} \ ,
\end{equation}
\begin{equation}
\tilde{\delta h_2} = \alpha K_1 \exp{(\lambda_1 t)} + \beta K_2 \exp{(\lambda_2 t)} \ ,
\end{equation}
\end{subequations}
where $(\lambda_1, \lambda_2)$  and $(1,K_1), (1,K_2)$ are the eigenvalues and eigenvectors of $\mathbf{s}$, respectively. The two coefficients $\alpha$ and $\beta$ can be found using the initial conditions: $\delta h_1(x,t=0) = 0$, and $\delta h_2(x,t = 0) = h_0 [\Theta(x) - 1/2] $, where $\Theta$ denotes the Heaviside function (\textit{i.e.} $\Theta(x>0) = 1$, $\Theta(x<0) = 0$). 

To evaluate and extend the validity of the \textit{asymptotic model} described so far, a second model was developed that does not assume any specific flow profiles in the two layers, and that takes into account all the terms of the Stokes equations, including the vertical velocities. We refer to this as the \emph{full-Stokes model} (see details in Appendix~\ref{AppendixStokes}). The \emph{full-Stokes model} exhibits governing equations similar to Eq.~\eqref{LinearizedGoverningEquation}, with the exception of the matrix elements $s^\textrm{Stokes}_{\textrm{i}, \textrm{j}}$, which are more complicated functions of $k$ than $s_{\textrm{i}, \textrm{j}}$. Excellent self-consistent agreement between the solutions of the two models is found in the small-slope limit (see  Appendix~\ref{AppendixStokes}).

\section{Results and Discussion}
\subsection{Interface profiles}
\label{section:profiles}
\begin{figure}[h]
\centering
\subfloat{\includegraphics[width = \textwidth]{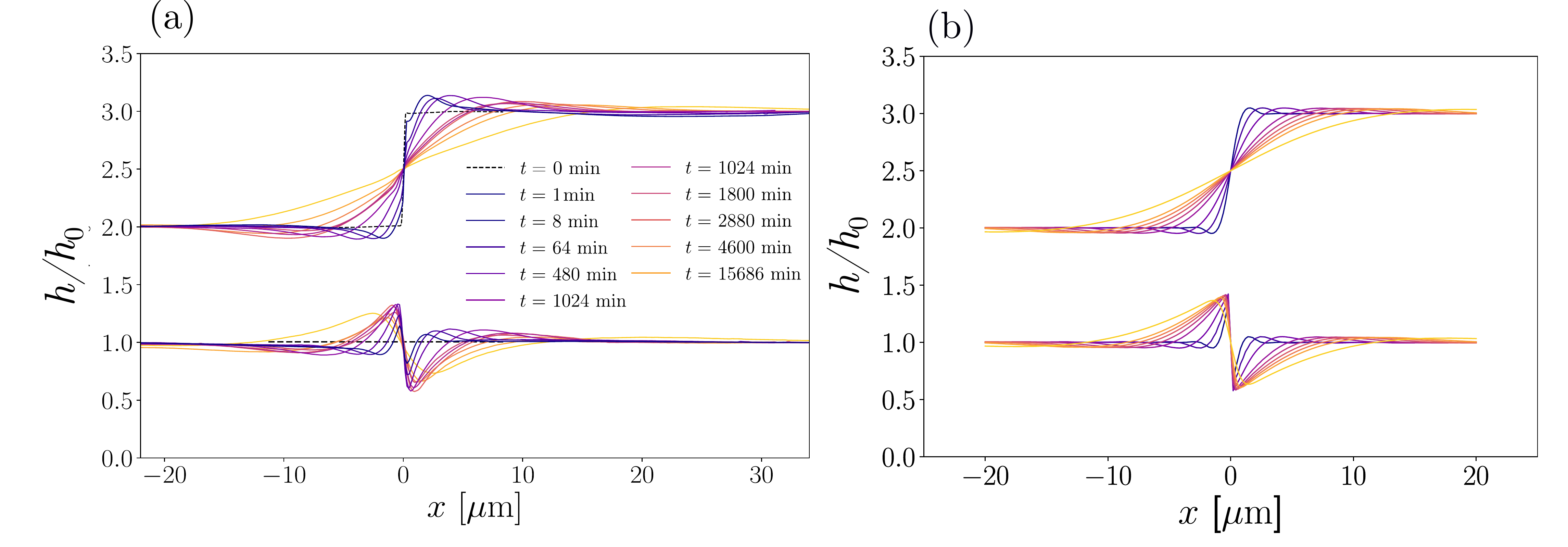}}
\caption{(a) Experimental profiles $h(x,t)=h_i(x,t)$ of the liquid-liquid ($i=1$, bottom) and liquid-air ($i=2$, top) interfaces of a PMMA ($M_{\textrm{w}}$ = 56~kg/mol) stepped layer on a PS ($M_{\textrm{w}}$ = 53.3~kg/mol) layer (see Fig.~\ref{fig:initcond}), during levelling at $T$ = 150 $^{\circ}$C. The viscosity ratio for these samples is $\mu = 1.1 \times 10^{-4}$ (see Tab.~\ref{table}). The samples were fabricated with $h_0 = 180$~nm. (b) Theoretical profiles calculated using the \emph{asymptotic model}. The times, rheological properties and geometry have been chosen to match the experimental conditions of the data shown in (a).}
\label{fig:profiles_topbot}
\end{figure}

In Fig.~\ref{fig:profiles_topbot}(a), we show the experimental profiles of the liquid-air and liquid-liquid interfaces at different stages of evolution for the case of PMMA with $M_{\textrm{w}} = 56$ kg/mol, PS with $M_{\textrm{w}} = 53$ kg/mol, and an annealing temperature of 150$^{\circ}$C. We note that each pair of interface profiles at a given annealing time corresponds to a different sample, as the top layer must be removed in order to image the buried liquid-liquid interface. Thus, a series of equivalent samples was prepared in order to reconstruct the entire evolution. Each sample was annealed for a given time, its liquid-air interface was imaged, the PMMA layer removed, and finally the profile of the bottom layer at the same location was imaged. The liquid-air interface develops a ``bump'' on the upper side of the step with positive curvature (\textit{i.e.} negative second derivative of the profile), and a ``dip'' on the lower side with negative curvature. With increasing annealing time, the bump and dip spread apart horizontally as the step levels. Furthermore, at late times, the bump and dip decrease in height. As discussed previously~\cite{SalezAnalytic}, the bump and dip develop to alleviate the large gradients in Laplace pressure due to the highly curved corners of the original stepped geometry. At early annealing times ($t < 8$ min), there is a sharp feature near the center of the step that seems to be a remnant of the initial corner of the step. 

The liquid-liquid interface deforms significantly in response to the Laplace pressure due to the stepped liquid-air interface. Remarkably, the deformation of the liquid-liquid interface initially grows vertically, before levelling out, which implies that while the surface energy associated with the liquid-air interface decreases, it partially does so at the cost of an increasing surface energy of the liquid-liquid interface. On either side of the deformation are shapes that mimic the bump and dip of the liquid-air interface. The deformation of the liquid-liquid interface can be qualitatively understood by considering the interfacial tension ratio $\Gamma=\gamma_1/\gamma_2$, as well as the viscosity ratio $\mu=\eta_1/\eta_2$ (see Sec.~\ref{subsec:mu} for a detailed study of the latter) introduced above. Since $\Gamma\ll1$, the liquid-liquid interface is much more compliant than the liquid-air interface, and hence the liquid-liquid interface adapts and follows the liquid-air interface. Moreover, the total interfacial energy of the system is dominated by the liquid-air contribution, as demonstrated quantitatively below (see Sec.~\ref{subsection:Energy}).

Figure~\ref{fig:profiles_topbot}(b) shows the theoretical profiles generated from the \emph{asymptotic model}, with all the physical parameters matching the experimental conditions of the data in Fig.~\ref{fig:profiles_topbot}(a). The model captures the essential features observed in the experiments, with the exception of a few early-time features (\textit{e.g.} initial vertical growth of the liquid-liquid interface and sharp feature near the step corner). In fact, at early times, the small-slope approximation is violated since $\frac{\partial h_2}{\partial x} \mid_{x=0}$ is of order one. We thus suggest that vertical flows, neglected in the \emph{asymptotic model}, are responsible for such features. The \emph{full-Stokes model}, which accounts for vertical flows, does capture these early-time details (see Appendix~\ref{AppendixStokes}), which supports our suggestion. 

\subsection{Effect of the viscosity ratio}
\label{subsec:mu}
\begin{figure}[t]
\centering
\subfloat{\includegraphics[width = \textwidth]{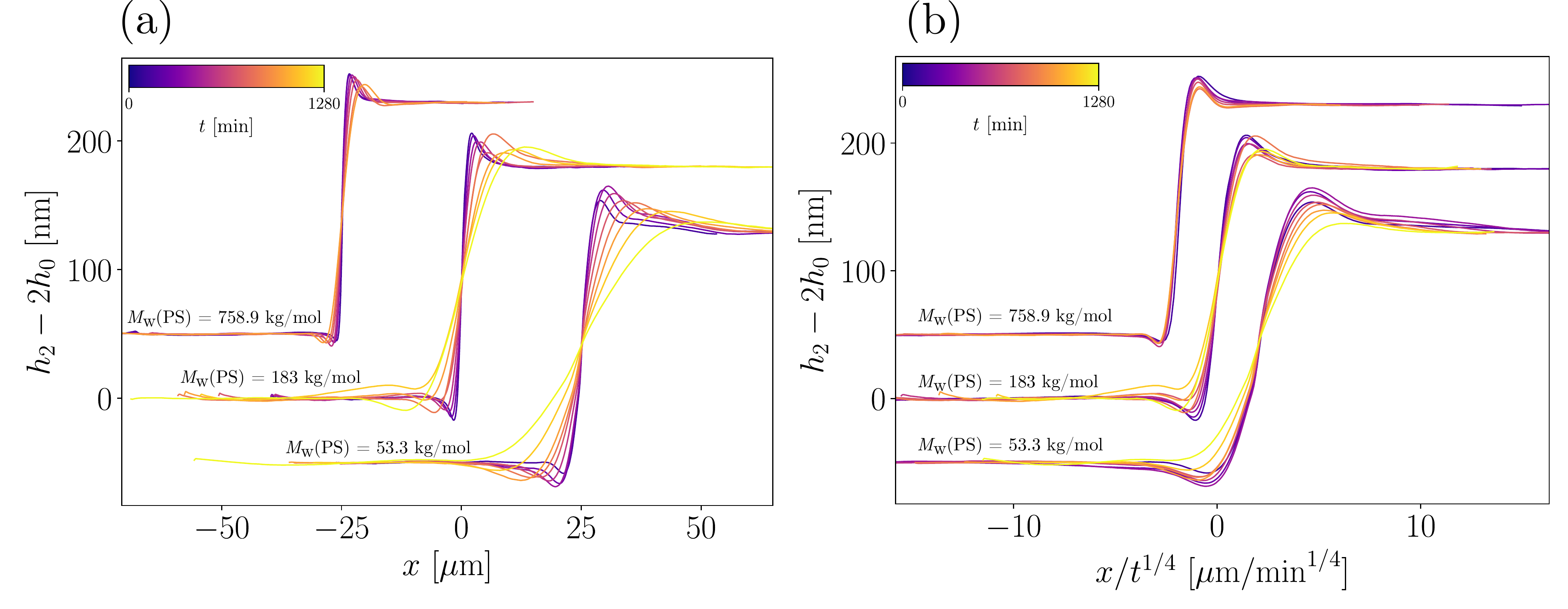}}
\caption{(a) Experimental liquid-air (\textit{i.e.} PMMA-air) interface profiles, at an annealing temperature of 165 $^{\circ}$C, for various annealing times and PS molecular weights, as indicated. (b) Same experimental data as in (a), but with a rescaled horizontal axis. For both panels, the 53.3 and 758.9 kg/mol data have been shifted horizontally and vertically for clarity.}
\label{fig:profiles_tops}
\end{figure}
Figure~\ref{fig:profiles_tops}(a) shows the experimental liquid-air (\textit{i.e.} PMMA-air) interface profiles, at an annealing temperature of 165~$^{\circ}$C, for various annealing times and PS molecular weights. For the experimental profiles in Figs.~\ref{fig:profiles_topbot}(a) and~\ref{fig:profiles_tops}, $\mu$ is always much smaller than 1 -- except in the case of $M_{\textrm{w}}(\textrm{PS})=758.9$~kg/mol at 165~$^{\circ}$C, for which $\mu$ is of order unity (see Tab.~\ref{table}). As explained in Sec.~\ref{section:profiles}, the evolution of the system is mostly driven by the gradients in Laplace pressure along the liquid-air interface. The resulting pressure field in the top PMMA layer is transferred to the underlying PS layer, thereby inducing flow in the latter and thus deformation of the liquid-liquid interface. Finally, it is immediately clear from Fig.~\ref{fig:profiles_tops}(a) that for samples having identical annealing temperatures, annealing times and geometry, the lower the viscosity of the underlying PS layer, the faster the levelling of the liquid-air interface. This highlights the importance of the bottom layer in the relaxation of the top layer, and is in line with the discussion in Sec.~\ref{subsec:theory} about the apparent slip length $b \sim h_1 / \mu$ in our configuration~\cite{jachalski2015thin}.

As discussed in the introduction, the capillary levelling of thin liquid films can exhibit self-similar regimes. For films supported on no-slip substrates and with the associated Poiseuille flow, the self-similar variable is $x/t^{1/4}$, while for freestanding films and plug flow, $x/t^{1/2}$ provides the appropriate rescaling. Figure~\ref{fig:profiles_tops}(b) shows the same data as in Fig.~\ref{fig:profiles_tops}(a) with the horizontal axis rescaled as expected for a  Poiseuille flow. For the largest viscosity ratio, obtained with $M_{\textrm{w}}(\textrm{PS})$ = 758.9 kg/mol, the rescaled profiles collapse well with one another. This is consistent with the physical intuition that a high enough viscosity in the bottom layer renders the situation analogous to capillary levelling on a solid substrate. However, for the two smaller viscosity ratios, there is no such collapse, which suggests that there is no $\sim t^{1/4}$ self-similar behaviour within the experimental temporal range. Similarly, rescaling the $x$-axis by $t^{1/2}$ (not shown) does not allow us to collapse the experimental profiles either. Therefore, in order to investigate the temporal evolution in more detail, we consider in Sec.~\ref{subsection:Energy} the evolution of the surface energy of the system -- \textit{i.e.} a global observable linked to capillary levelling~\cite{McGrawEnergy}. 

\subsection{Energetic considerations}
\label{subsection:Energy}
\begin{figure}
\centering
\subfloat{\includegraphics[width = 0.6\textwidth]{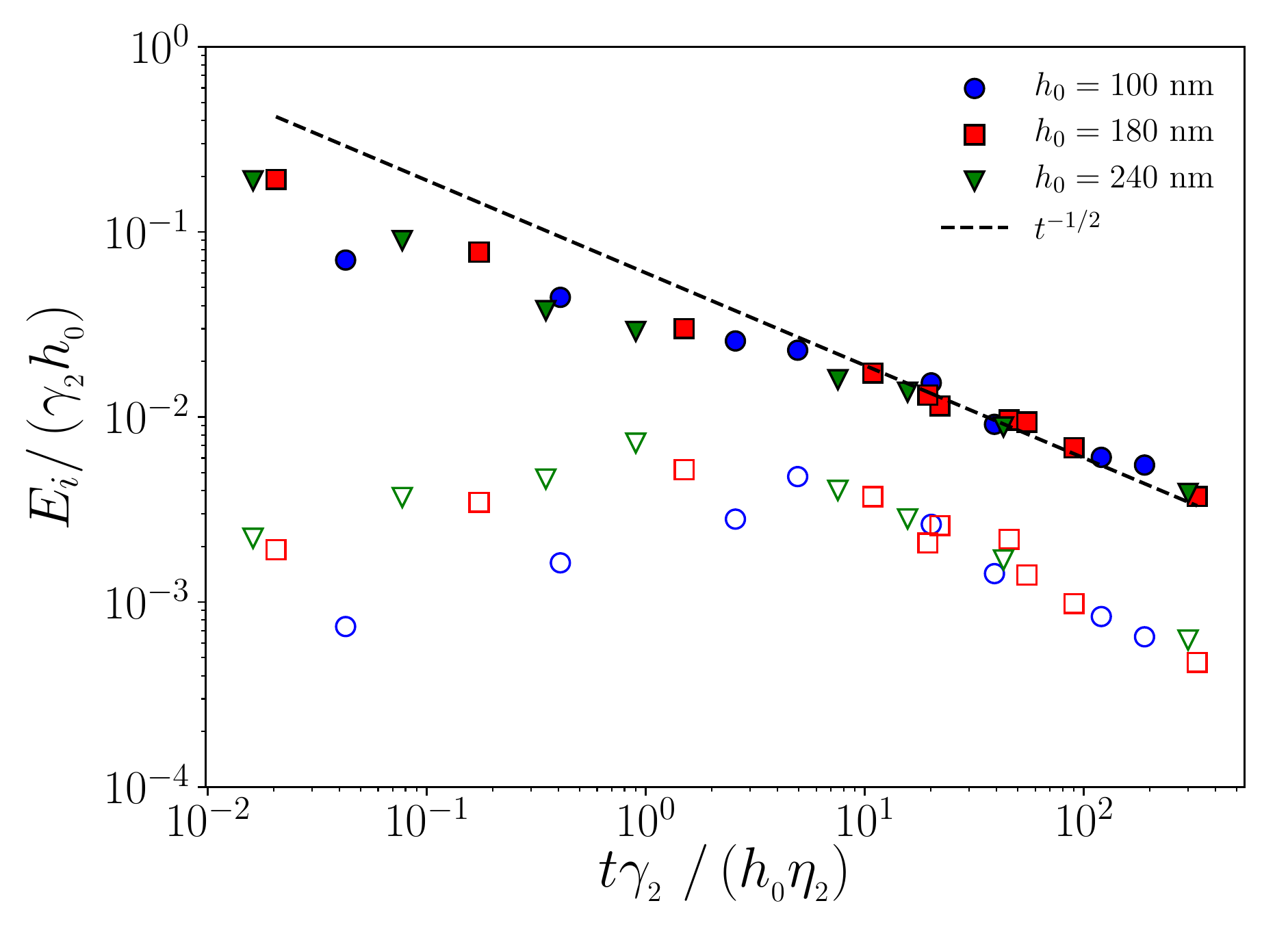}}
\caption{Normalized excess capillary energy per unit length of the liquid-liquid ($i=1$, unfilled) and liquid-air ($i=2$, filled) interfaces, for PMMA ($M_\textrm{w}$ = 56 kg/mol) stepped films on PS ($M_\textrm{w}$ = 53.3 kg/mol) films, with three different values of the nominal thickness $h_0$ (see Fig.~\ref{fig:initcond}) as indicated, and an annealing temperature of 150 $^{\circ}$C. The excess capillary energies per unit length have been normalized by the corresponding initial values for the liquid-air interface. Longterm $\sim t^{-1/2}$ behaviours are indicated with dashed lines.}
\label{fig:energy_topbot}
\end{figure}

The excess capillary energy $\mathcal{E}_i$ of interface $i$ is proportional to the interfacial tension $\gamma_i$, as well as to the difference between the interfacial area $S_i$ and the interfacial area $S_i^{\,0}$ of the flat equilibrium state: $\mathcal{E}_i = \gamma_i (S_i - S_i^{\,0})$. Given the invariance of the system with respect to the $y$-direction, and relating the interfacial lengths to the local profiles $h_i(x)$, we consider the excess capillary energies per unit length defined as: $E_i = \gamma_i \int  \textrm{d}x\, (\sqrt{1 + h_i'(x)^2} - 1)$. 
In order to account for the different initial liquid-liquid interfacial lengths, resulting from the different $h_0$ values and thus step heights, the excess capillary energies per unit length can be normalized by the corresponding initial values $\gamma_2 h_0$ for the liquid-air interface. 
In Fig.~\ref{fig:energy_topbot} the normalized excess capillary energy per unit length is plotted versus dimensionless time for both the liquid-liquid ($i=1$) and liquid-air ($i=2$) interfaces, from the data shown in Fig.~\ref{fig:profiles_topbot}(a), as well as from data obtained with two other thicknesses $h_0$. 
At dimensionless times $t\gamma_2/(h_0\eta )$ larger than $\sim10^4$, one observes that the excess capillary energies of both interfaces seem to decrease as $t^{-1/2}$ power laws. In addition, as expected and discussed in Sec.~\ref{section:profiles}, the contribution of the liquid-liquid interface is $\sim7$ times smaller than that of the liquid-air interface. This dominance of the liquid-air excess interfacial energy to the total interfacial energy confirms the intuitive remark made previously that the liquid-liquid interface deforms with a relatively little cost since $\Gamma \ll 1$. In the following, we thus focus on the liquid-air interface alone. Conveniently, one can then prepare a single sample and follow the evolution of the liquid-air interface through repeated annealing, rather than having to sacrifice the sample by dissolving the top PMMA layer.
The excess capillary energies per unit length can be computed from the \emph{asymptotic model}, \textit{\textit{i.e.}} in the small-slope limit, with the approximation $E_i \simeq \gamma_i \int \textrm{d}x\, h_i'(x)^2/2$. We note that, under this approximation, if the profile of a given interface is self-similar, such that $h_i(x,t) = f_i(x/t^\alpha)$ with $f_i$ a function of a single variable, then $E_i\sim t^{-\alpha}$. Therefore, $E_i\sim t^{-1/4}$ or $E_i\sim t^{-1/2}$ indicate the dominance of Poiseuille or plug flows, respectively, as discussed in Sec.~\ref{subsec:mu}. 

\begin{figure}[t]
\centering
\includegraphics[width = 1\textwidth]{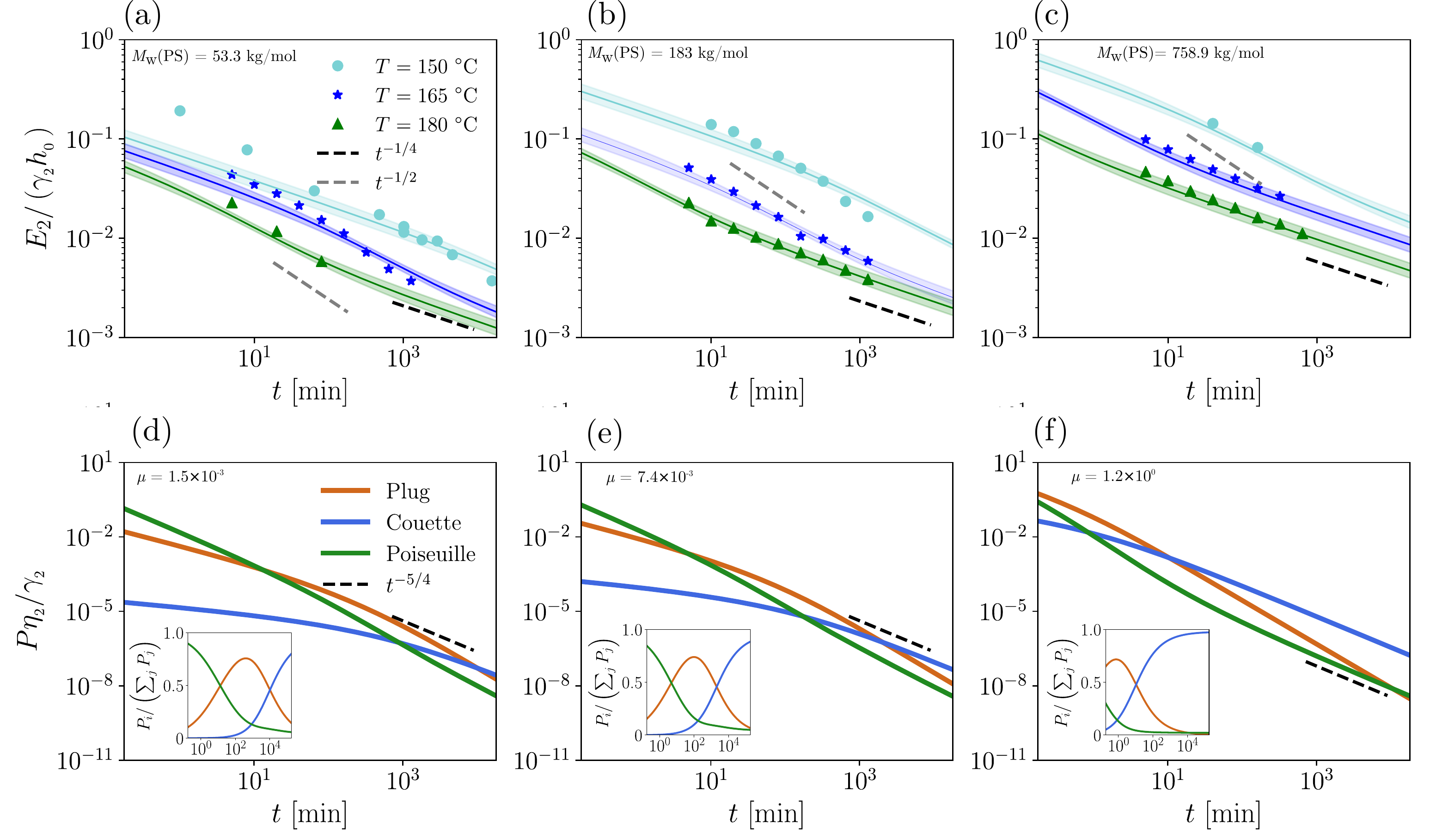}
\caption{(a) - (c) Normalized excess capillary energy per unit length $E_2/(\gamma_2h_0)$ of the liquid-air interface as a function of time $t$, for PMMA ($M_{\textrm{w}}$ = 56 kg/mol) stepped layers on PS layers, annealed at 150 $^{\circ}$C ($\circ$), 165$^{\circ}$C ($\star$) and 180 $^{\circ}$C ($\triangle$). The PS molecular weights are: (a) $M_\textrm{w}= 53.3$~kg/mol, (b) $M_\textrm{w}= 183$~kg/mol, and (c) $M_\textrm{w} = 758.9$~kg/mol. The solid lines of corresponding colors indicate the normalized excess capillary energy per unit length $E_2/(\gamma_2h_0)$ of the top interface calculated from the \textit{asymptotic model}, with the PS viscosity as a single fitting parameter. The bands span a variation in viscosity ratio from half of the best fit to double of the best fit. The excess capillary energies per unit length have been normalized by the corresponding initial values for the liquid-air interface. The $t^{-1/2}$ and $t^{-1/4}$ trends are indicated with dashed lines. (d) - (f) Normalized viscous dissipation powers per unit length $P\eta_2/\gamma_2$ as a function of time $t$, calculated from the \textit{asymptotic model}, for the three viscous mechanisms of Eq.~\eqref{Equation:EnergyBalance}: plug ($P=P_{\textrm{plug}}$), Poiseuille ($P=P_{\textrm{Poiseuille}}$) and Couette ($P=P_{\textrm{Couette}}$). For each panel, the viscosity ratio $\mu$ corresponds to the one of the 165 $^{\circ}$C data in the panel above. The $t^{-5/4}$ trend is indicated with dashed lines. The insets display the respective fractions of the total viscous dissipation on the same time scale.}
\label{fig:energy}
\end{figure}

Figures~\ref{fig:energy}(a)-(c) show the normalized excess capillary energy per unit length $E_2$ of the liquid-air interface as a function of time $t$, for different annealing temperatures and PS molecular weights. For each panel, three identical samples were prepared and annealed at different temperatures. The experimental data are overlaid with best fits to the excess capillary energy per unit length $E_2$ of the liquid-air interface obtained from the \textit{asymptotic model}, using the PS viscosity as the single free parameter. We note an excellent agreement between experiments and theory, except at the earliest times for the sample made with a 53.3 kg/mol PS molecular weight and annealed at 150$^\circ$C. In that case, the experimental values are substantially higher than predicted by the model, which is likely due to the sharp feature observed at early times (see Fig.~\ref{fig:profiles_topbot}(a)), as noted in Sec.~\ref{section:profiles}. Indeed, this feature cannot be captured by the \textit{asymptotic model} (see Fig.~\ref{fig:profiles_topbot}(b)) which neglects any vertical flow (see Appendix~\ref{AppendixStokes}), and it would elevate the capillary energy compared to a profile without that feature. Finally, the viscosity ratios obtained from the fits are in good agreement with independent measurements (see Tab.~\ref{table}) -- both approaches being within an order of magnitude of each other. 

While the experimental data in Figs.~\ref{fig:energy_topbot} and~\ref{fig:energy}(a) seems consistent with a  $E_2\sim t^{-1/2}$ trend, the \textit{asymptotic model} does not predict a well-defined regime with such a scaling law. Instead, the model seems to exhibit a double crossover, with a transient exponent that changes smoothly from a value near $-1/4$, to a value near $-1/2$, before increasing again towards $-1/4$, as seen in Figs.~\ref{fig:energy}(a)-(c). The latter seems to correspond to a proper long-term self-similar regime, valid for all viscosity ratios. Thus, the energy becomes eventually independent of the viscosity $\eta_2$ of the top film, which indicates that most of the dissipation occurs in the bottom film at late times. This long-term self-similar regime is reached experimentally in some cases (see Figs.~\ref{fig:energy}(b)-(c)), but is not accessible for the smallest viscosity ratios due to the large experimental time scales involved. The apparent $E_2\sim t^{-1/2}$ regime observed in Figs.~\ref{fig:energy_topbot} and~\ref{fig:energy}(a) is thus a transient, intermediate behaviour. 

To gain further insight into the effect of the viscosity ratio, we express the conservation of energy within the \emph{asymptotic model}. As the films are thin, body forces may be neglected. The capillary energy decreases primarily through viscous dissipation. The rate of change of the total capillary energy per unit length can be written as the sum of the three viscous dissipation powers induced by the characteristic flows highlighted in Sec.~\ref{subsec:theory}:
\begin{equation}
\label{Equation:EnergyBalance}
\partial_t E = - \underbrace{\int \textrm{dx}\, 4\eta_2 \, (h_2 - h_1)  u_2'^{2}}_{P_{\textrm{plug}}}   - \underbrace{\int \textrm{dx}\, \frac{ p_1'^2 \, h_1^{\,3}}{12\eta_1}}_{P_{\textrm{Poiseuille}}} - \underbrace{\int \textrm{dx}\,  \eta_1 \frac{u_2^{\,2}}{h_1}}_{P_{\textrm{Couette}}}\ .
\end{equation}
An explicit derivation of Eq.~(\ref{Equation:EnergyBalance}) is provided in Appendix~\ref{AppendixEnergy}. The first term ($P_{\textrm{plug}}$) is related to a velocity profile that is invariant vertically through the top layer's thickness, corresponding to plug flow. The second term ($P_{\textrm{Poiseuille}}$) is related to a parabolic velocity profile, \textit{i.e.} Poiseuille flow, caused by the horizontal Laplace pressure gradient in the bottom layer. The third term ($P_{\textrm{Couette}}$) corresponds to a linear variation in the velocity profile of the bottom layer, as seen in a simple shear geometry or Couette flow.

Figures~\ref{fig:energy}(d)-(f) display the normalized viscous dissipation powers per unit length as a function of the rescaled time, calculated from the \textit{asymptotic model}, for the three viscous mechanisms of Eq.~\eqref{Equation:EnergyBalance}, and for three experimentally-relevant viscosity ratios. Three main regimes can be identified from the respective fractions of the total viscous dissipation power shown in the insets. At early times, the Poiseuille contribution dominates, which is associated with a $E\sim t^{-1/4}$ behaviour~\cite{McGrawEnergy} and thus $\partial_tE\sim t^{-5/4}$. At late times, the Couette contribution dominates, but since the Couette flow in the bottom layer is indirectly induced by the Laplace pressure gradient from the liquid-air interface, it also exhibits a $E\sim t^{-1/4}$ power law like the Poiseuille flow. Therefore, at late times, we also expect a $\partial_tE\sim t^{-5/4}$ behaviour. Furthermore, we recover the result stated above that the dissipation occurs mostly in the bottom film in this regime. Finally, at intermediate times, in between these two extreme regimes, the plug contribution seems to dominate. This is associated with a transient temporal exponent for the energy, passing by the $-1/2$ value~\cite{IltonFreestanding}. All together, we recover the non-monotonic trend for the temporal exponent discussed above from the theoretical predictions in Figs~\ref{fig:energy}(a)-(c), and we can further characterize it as a Poiseuille-to-plug-to-Couette double crossover.

In the case where $\mu \sim 1$, valid for PS with a molecular weight of 758.9 kg/mol, the shear stress at the liquid-liquid interface does not vanish and therefore shear terms must be taken into account in the top layer. The \textit{asymptotic model} should thus be refined for bilayer films with such material properties (see \textit{lubrication model} in Appendix~\ref{Appendix_LargeViscosity}). However, according to the \textit{asymptotic model}, the late-time dissipation is mainly dominated by the Couette contribution in the bottom layer (see Fig.~\ref{fig:energy}(f)), and this model reproduces qualitatively the data and in particular the $x/t^{1/4}$ self-similarity of the profiles (see Fig.~\ref{fig:profiles_tops}(b)). Nevertheless, we stress that the prefactor of the late-time scaling law $E_2 \sim t^{-1/4}$ from the \textit{asymptotic model} is larger than the one from the \textit{lubrication model}, as observed in Fig.~\ref{fig:energy_StokesAsymptotic} (see Appendix~\ref{Appendix_LargeViscosity}). As a result, the fitting procedure leads to a systematic underestimation of the viscosity of the bottom layer, as confirmed in Tab.~\ref{table}. 

Finally, as stated in Sec.~\ref{subsec:theory}, the set of equations (see Eqs.~(\ref{GovEq1}),~(\ref{GovEq2}) and~(\ref{GovEq3})) that form the \emph{asymptotic model} are reminiscent of the equations that describe capillary levelling on a substrate with slip~\cite{munch2005lubrication}. For a nearly-constant bottom layer thickness $h_1$, the Couette dissipation power per unit length $P_{\textrm{plug}}=\eta_1 \int \textrm{d}x\, u_2^{\,2}/h_1$ (see Eq.~(\ref{Equation:EnergyBalance})) is indeed similar to the power per unit length $k \int \textrm{d}x\, u_2^{\,2}$ dissipated on a solid substrate through friction, provided that the friction coefficient $k$ is identified to $\eta_1/h_1$ and the slip velocity to $u_2$. As a consequence, the bottom film acts as a lubrication layer below the top stepped film, which leads to an apparent slip length given by $b \sim h_1 \eta_2/\eta_1$. As such, our first (\textit{i.e.} Poiseuille-to-plug) crossover mimics the one expected for a single film supported on a rigid substrate with varying slip boundary condition~\cite{munch2005lubrication,McGraw2016}.

\section{Conclusion}
In this article, we examined the effect of a thin liquid substrate on the capillary levelling of a thin liquid film placed atop. Specifically, we prepared stepped polymer layers that were placed onto other, immiscible and flat polymer layers supported on solid substrates. The bilayer films were observed to flow and relax towards a flat equilibrium configuration. We showed that the liquid-liquid interface deforms substantially. In the samples studied, the viscosity ratio between the two layers was varied over a large range, with the bottom layer being less viscous, or as viscous as the top layer. We have shown that the viscosity ratio has a major impact on the resulting dynamics. Unlike the capillary levelling of simple stepped films on solid substrates, or freestanding films, the interfacial profiles do not exhibit any clear, unique and stable self-similar behaviour. We have developed and validated a thin-film model in which the governing flow in the top layer is plug-like, and flow in the bottom layer is a with a combination of Poiseuille and Couette flows. Using an energetic treatment, we have shown that the excess capillary energy introduced by the step, with respect to the flat equilibrium state, is dissipated by those three coupled viscous mechanisms, thus leading to a novel Poiseuille-to-plug-to-Couette double crossover. The time scales in the process depend on the viscosity ratio between the bottom and top layers. We have found that the bottom, less viscous layer is analogous to a solid substrate with a certain finite slip length. The experimentally-measured energy dissipation is in agreement with that obtained from the model. The results presented illuminate the intricate dynamics of viscous bilayer assemblies, and might find applications through friction control by lubrication, self-assembly and stability of multilayer processes.
 
\section*{Acknowledgments}
The authors are grateful to John Niven for valuable insight and discussions. We gratefully acknowledge financial support by the Natural Science and Engineering Research Council (NSERC) of Canada. 

\appendix 
\section{Asymptotic model}
\label{AppendixAsymptotic}
\subsection{Model}
This appendix expands upon the \emph{asymptotic model} discussed in Sec.~\ref{subsec:theory}. Dimensionless variables are denoted by capital letters:
\begin{equation}
u_i = u \, U_i, \quad w_i = w W_i = u \epsilon W_i, \quad x = l \, X, \quad z = h_0 \, Z, \quad p_i = p P_i, \quad t = \frac{l}{u} \, T, \quad h_i = h_0 \, H_i, \quad \epsilon = \frac{h_0}{l},
\end{equation}
where $\epsilon$ is the ratio between the typical vertical scale $h_0$ (see Fig.~\ref{fig:initcond}) and an horizontal length scale $l$, $P = \frac{\gamma_2 h_0}{l^2}$ is the typical pressure scale set by the Laplace pressure, and $u = \frac{\gamma_2 h_0}{\eta_2 l}$ is the characteristic velocity which is chosen such that the leading-order equation for the top layer is compatible with plug flow. We note that there is no intrinsic horizontal length scale in our configuration, due to the initial stepped geometry. Therefore, $l$ should be estimated as a typical width of the levelling profile~\citep{McGrawViscosity}. Thus, the \emph{asymptotic model} is valid when this length scale is much larger than the typical height $h_0$. We rescale the viscosity ratio as $M = \epsilon^{-2} \mu$ for appropriate governing equations in the bottom layer~\cite{jachalski2015thin}. Nondimensionalization yields the governing Stokes equations for both viscous layers:
\begin{subequations}
\begin{equation}
0 = - \epsilon^2 \, \partial_X P_2 + \epsilon^2 \partial^2_X U_2 + \partial_Z^2 U_2,
\end{equation}
\begin{equation}
0 = -  \, \partial_Z P_2 + \epsilon^2 \partial^2_X W_2 +  \partial_Z^2 W_2,
\end{equation}
\begin{equation}
\partial_X U_2 + \partial_Z W_2  = 0,
\end{equation}
\begin{equation}
0 = - \, \partial_X P_1 + M (\epsilon^2 \partial^2_X U_2 + \partial_Z^2 U_1),
\end{equation}
\begin{equation}
0 = -  \, \partial_Z P_1 + M (\epsilon^4 \partial^2_X W_1 + \epsilon^2 \partial_Z^2 W_1),
\end{equation}
\begin{equation}
\partial_X U_1 + \partial_Z W_1  = 0.
\end{equation}
\end{subequations}
At the free interface, the boundary conditions are the stress balance and the kinematic condition~\cite{jachalski2014}:
\begin{subequations}
\begin{equation}
P_2 + \frac{\partial^2_X H_2}{[1 + \epsilon^2(\partial_X H_2)^2]^{3/2}} = 2 \frac{ \partial_Z W_2 [1 - \epsilon^2(\partial_X H_2)^2] - (\partial_Z U_2 + \epsilon^2 \partial_X W_2 )\partial_X H_2}{1 + \epsilon^2(\partial_X H_2)^2} , \quad \quad Z = H_2,
\end{equation}
\begin{equation}
\label{TangStress2}
(\partial_Z U_2 + \epsilon^2 \partial_X W_2)[1 - \epsilon^2 (\partial_X H_2)^2] = 4 \epsilon^2 \partial_X U_2 \partial_X H_2 , \quad \quad Z = H_2,
\end{equation}
\begin{equation}
\partial_T H_2 = W_2 - U_2 \partial_X H_2 , \quad \quad z = h_2.
\end{equation}
\end{subequations}
The boundary conditions at the liquid-liquid interface are the stress balance and the kinematic condition. Furthermore, we assume that there is no slip at the interface. All together, these read:
\begin{subequations}
\begin{equation}
\begin{split}
P_1 - P_2 + \Gamma \frac{\partial^2_X H_1}{[1 + \epsilon^2(\partial_X H_1)^2]^{3/2}} = \, & 2 \frac{ \partial_Z (M \epsilon^2 W_1 - W_2) [1 - \epsilon^2(\partial_X H_1)^2] }{1 + \epsilon^2(\partial_X H_1)^2}  \\ 
&- 2\frac{[\partial_Z (M \epsilon^2 U_1 - U_2) + \epsilon^2 \partial_X (M \epsilon^2 W_1 - W_2) ]\partial_X H_1 }{1 + \epsilon^2(\partial_X H_1)^2}  , \quad \quad Z = H_1
\end{split}
\end{equation}
\begin{equation}
\label{TangStress1}
\left[\partial_Z (M \epsilon^2 U_1 - U_2) + \epsilon^2 \partial_X (M \epsilon^2 W_1 - W_2) \right] \left[1 - \epsilon^2 (\partial_X H_2)^2\right]  = 4 \epsilon^2 \partial_X (M \epsilon^2 U_1 - U_2) \partial_X H_1  , \quad \quad Z = H_1
\end{equation}
\begin{equation}
\partial_T H_1 = W_1 - \partial_X H_1 , \quad \quad Z = H_1
\end{equation}
\begin{equation}
W_2 -  W_1 = (U_2 -  U_1)\partial_X H_1 , \quad \quad Z = H_1
\end{equation}
\begin{equation}
U_2 -  U_1 + \epsilon^2 (W_2 -  W_1) \partial_X H_1 = 0  , \quad \quad Z = H_1.
\end{equation}
\end{subequations}
At the solid-liquid interface, we assume a no-slip boundary condition:
\begin{equation}
U_1 = W_1 = 0 , \quad \quad Z = 0.
\end{equation}
We consider the flow in the top layer as a perturbation:
\begin{equation}
\left(U_2, W_2, P_2\right) = \left( U_2^{(0)}, W_2^{(0)}, P_2^{(0)}\right) + \epsilon^2 \left(U_2^{(1)}, W_2^{(1)}, P_2^{(1)}\right).
\end{equation}
The leading order can be described as:
\begin{subequations}
\begin{equation}
\partial_Z U_2^{(0)}(X,Z,T) = 0 \quad \rightarrow \quad U_2^{(0)}(X,Z,T) = U_2(X,T) ,
\end{equation}
\begin{equation}
W_2^{(0)}(X,Z,T) = - (Z-H_1)\partial_X U_2 + W_1(Z=H_1),
\end{equation}
\begin{equation}
P_2^{(0)}(X,T) = - 2 \partial_X U_2 - \partial^2_X H_2,
\end{equation}
\begin{equation}
\partial_Z P_1(X,Z,T) = 0 \quad \rightarrow \quad P_1(X,T) = -\partial^2_X H_2(X,T) - \Gamma \partial^2_X H_1(X,T)
\end{equation}
\begin{equation}
U_1(X, Z, T) = -\frac{1}{2M} \partial_X P_1 \left( Z^2 - ZH_1 \right) + U_2(X,T) \frac{Z}{H_1}.
\end{equation}
\end{subequations}
The in-plane component of the flow is described by a set of coupled non-linear equations. Invoking further the kinematic condition results in Eq.\eqref{GovEq1}:
\begin{equation}
\partial_t (H_2 - H_1) = - \left[ (H_2 - H_1)U_2 \right]'\, ,
\end{equation}
where the prime denotes the derivative with respect to $X$. The volume conservation of the bottom layer gives: Eq.~\eqref{GovEq2}:
\begin{equation}
\partial_t H_1 = -\left(-P_1'\frac{H_1^3}{12M} +\frac{H_1 U_2}{2}   \right)' = -\left[( H_2'''+\Gamma H_1''')\frac{H_1^3}{12M} +\frac{H_1 U_2}{2}   \right]'.
\end{equation}
The final equation that relates $U_2$ to the other variables can be found by integrating the horizontal component of the Stokes equation with respect to $Z$ at the next leading order:
\begin{equation}
 \partial^2_Z U_2^{(1)} + \partial_X^2 U_2^{(0)} = \partial_X P_2^{(0)} \quad \rightarrow \quad \partial_Z U_2^{(1)} (Z = H_2) - \partial_Z U_2^{(1)}(Z=H_1) = \left( \partial_X P_2^{(0)} - \partial^2_X U_2 \right)(H_2 - H_1).
\end{equation}
We find the last governing equation, Eq.~\eqref{GovEq3}, by inserting the two tangential stress balances, Eq.~\eqref{TangStress2} and Eq.\eqref{TangStress1}, at leading order into the previous equation:
 \begin{equation}
H_2''' (H_2 - H_1) + (H_2''' + \Gamma H_1''') H_1/2  + 4[U_2' (H_2-H_1)]' - M \frac{U_2}{H_1} = 0 .
\end{equation}
\subsection{Decay rates}
The elements $s_{i,j}$ of the decay-rate matrix $\mathbf{s}$ are found by taking the Fourier transforms of the linearized governing equations:
\begin{subequations}
\begin{equation}
s_{1,1}(k) = - \gamma_1 k^4 \left[ \frac{\bar{h}_1^3}{12\eta_1} + \frac{\bar{h}_1^3}{4(\eta_1 + 4\eta_2 k^2\bar{h}_1\bar{h}_2)} \right], \quad \quad s_{1,2}(k) = - \gamma_2 k^4 \left[ \frac{\bar{h}_1^3}{12\eta_1} + \frac{\bar{h}_1^2\bar{h}_2(1 - \frac{\bar{h}_1}{2\bar{h}_2})}{2(\eta_1 + 4\eta_2 k^2\bar{h}_1\bar{h}_2)} \right],
\end{equation}
\begin{equation}
s_{2,1}(k) = - \gamma_1 k^4 \left[ \frac{\bar{h}_1^3}{12\eta_1} + \frac{\bar{h}_1^2\bar{h}_2(1 - \frac{\bar{h}_1}{2\bar{h}_2})}{2(\eta_1 + 4\eta_2 k^2\bar{h}_1\bar{h}_2)} \right], \quad \quad s_{2,2}(k) = - \gamma_2 k^4 \left[ \frac{\bar{h}_1^3}{12\eta_1} + \frac{\bar{h}_1 \bar{h}_2^2 (1 - \frac{\bar{h}_1}{2\bar{h}_2})^2 }{(\eta_1 + 4\eta_2 k^2\bar{h}_1\bar{h}_2)} \right].
\end{equation}
\end{subequations}
The eigenvalues are the decay rates, and are given by:
\begin{equation}
\lambda_i = \frac{\textrm{Tr}(\mathbf{s}) \pm \sqrt{\textrm{Tr}(\mathbf{s})^2 - 4 \, \textrm{Det}(\mathbf{s}) }}{2}.
\end{equation}
The eigenvectors of $\mathbf{s}$ take the form $(1, K_i)$ with:
\begin{equation}
K_i = \frac{\lambda_i - s_{1,1}}{s_{1,2}}.
\end{equation}

\section{Full-Stokes model}
\label{AppendixStokes}
\subsection{Model}
To remove any assumption associated with pre-supposed flow types and the temporal scalings of the capillary energies, we derive a model from the Stokes equations~\cite{huang2002instability,Rivetti2017}. The stream functions $\psi_i$ of each layer ($i = 1, 2$) are defined as:
\begin{subequations}
\begin{equation}
u_i = -\partial_z \psi_i,
\end{equation}
\begin{equation}
w_i = \partial_x \psi_i.
\end{equation}
\end{subequations}
The velocity fields satisfy the Stokes equations. This in turn implies that the stream functions are solutions of biharmonic equations:
\begin{equation}
(\partial^4_x  +2 \partial_x^2\partial_z^2 + \partial_z^4) \psi_i = 0.
\end{equation}
We take the Fourier transforms (defined in the main text) with respect to the variable $x$, of the biharmonic equations, through the Fourier transforms $\tilde{\psi}_i$ of the stream functions, which results in fourth-order ordinary differential equations:
\begin{equation}
\left(\frac{d}{dz}\right)^4 \tilde{\psi_i} - \left(\frac{d}{dz}\right)^2k^2\tilde{\psi_i} + k^4 \tilde{\psi_i} = 0.
\end{equation}
The  general solutions are:
\begin{equation}
\tilde{\psi_i} (k,z) = A_\textrm{i}(k) \cosh(kz) + B_\textrm{i}(k) \sinh(kz)+C_\textrm{i}(k) z\cosh(kz)+D_\textrm{i}(k) z\sinh(kz).
\end{equation}
The eight coefficients $A_i, B_i, C_i, D_i$ can be found using the boundary conditions: vanishing velocity at the solid-liquid interface, continuity of velocity (including no slip) and stress across the liquid-liquid interface, and continuity of stress (including no shear stress) at the liquid-air interface. The non-linear terms of the curvature in the Laplace pressure are neglected, as well as the non-linear terms of the normal and tangential vectors to the interfaces. This means that this model would be valid in the limit of small slopes. The boundary condition are listed below:
\begin{subequations}
\begin{equation}
w_1 = 0 \quad \rightarrow \quad  \tilde{\psi_1} = 0 , \quad  z = 0,
\end{equation}
\begin{equation}
u_1 = 0 \quad \rightarrow \quad  \left(\frac{d}{dz}\right) \tilde{\psi_1} = 0,  \quad  z = 0,
\end{equation}
\begin{equation}
w_2 = w_1 \quad \rightarrow \quad  -ik\tilde{\psi_2}= -ik\tilde{\psi_1} , \quad  z = h_1,
\end{equation}
\begin{equation}
u_2 = u_1  \quad \rightarrow \quad  -\left(\frac{d}{dz}\right)\tilde{\psi_1} = -\left(\frac{d}{dz}\right)\tilde{\psi_2}, \quad  z = h_1,
\end{equation}
\begin{equation}
\eta_2 (\partial_z  u_2 + \partial_x  w_2) = \eta_1 (\partial_z u_1 + \partial_x  w_1) \quad \rightarrow \quad  \eta_1 \left[ \left(\frac{d}{dz}\right)^2\psi_1 + k^2\psi_1 \right] = \eta_2\left[\left(\frac{d}{dz}\right)^2\psi_2 + k^2\psi_2 \right], \quad  z = h_1.
\end{equation}
\begin{equation}
\begin{array}{c}
   - (p_1- p_2) + 2 \partial_z (\eta_1 w_1 - \eta_2 w_2) = -\gamma_1 \partial_x^2 h_1 \quad \rightarrow \quad     \\ \\
      \eta_1 \left[3 k^2 \left(\frac{d}{dz}\right)\tilde{\psi_1} - \left(\frac{d}{dz}\right)^3\tilde{\psi_1} \right] - \eta_2 \left[3 k^2 \left(\frac{d}{dz}\right)\tilde{\psi_2} - \left(\frac{d}{dz}\right)^3\tilde{\psi_2} \right] =  ik^3 \gamma_1 \tilde{h_1}, \quad  z = h_1.
\end{array}
\end{equation}
\begin{equation}
\eta_2 (\partial_z u_2 + \partial_x w_2) = 0 \quad \rightarrow \quad \left(\frac{d}{dz}\right)^2 \psi_2 + k^2\psi_2 = 0, \quad \quad z= h_2.
\end{equation}
\begin{equation}
-p_2 + 2\eta_2 \partial_z w_2 = -\gamma_2 \partial_x^2 h_2 \quad \rightarrow \quad  \eta_2 \left[3 k^2 \left(\frac{d}{dz}\right) \tilde{\psi_2} - \left(\frac{d}{dz}\right)^3\tilde{\psi_2} \right] = ik^3 \gamma_2\tilde{h_2},  \quad \quad z = h_2.
\end{equation}
\end{subequations}
The Stokes equations in the $x$-direction read:
\begin{equation}
\partial_x p_i = \eta_i (\partial^2_x u_i + \partial^2_z u_i) \quad \rightarrow \quad  -ik\tilde{p_i} = \eta_i \left[k^2\frac{d}{dz}\tilde{\psi_i} - \left(\frac{d}{dz}\right)\tilde{\psi_i}'''\right].
\end{equation}
The governing equations for the temporal evolutions of the thickness profiles can be found using the kinematic conditions: 
\begin{equation}
\partial_t h_i + u_i \partial_x h_i  = w_i,
\end{equation}
where $u_i$ and $w_i$ are evaluated at $z = h_i$. We further invoke small interfacial perturbations and proceed to linearization as in the \textit{asymptotic model}:
\begin{subequations}
\begin{equation}
\partial_t \tilde{\delta h_1} = \tilde{w_1}(z = \bar{h}_1) = s^\textrm{Stokes}_{11} \tilde{\delta h_1} + s^\textrm{Stokes}_{12} \tilde{\delta h_2}\, ,
\end{equation}
\begin{equation}
\partial_t \tilde{\delta h_2} = \tilde{w_2}(z = \bar{h}_1+\bar{h}_2) = s^\textrm{Stokes}_{21} \tilde{\delta h_1} + s^\textrm{Stokes}_{22} \tilde{\delta h_2}\, .
\end{equation}
\end{subequations}
These equations have the same general solutions as in the \emph{asymptotic model} developed in Sec.~\ref{subsec:theory}. The elements $s_{i,j}^\textrm{Stokes}$ of the decay-rate matrix are not written here but can be found using a formal calculation software. It is then straightforward to write the solutions as in Eq.~\eqref{eq:solution} with the corresponding eigenvalues $\lambda_i^\textrm{Stokes}$ and eigenvectors $(1, K_i^\textrm{Stokes})$.
\begin{figure}[h]
\centering
\includegraphics[width = 0.5\textwidth]{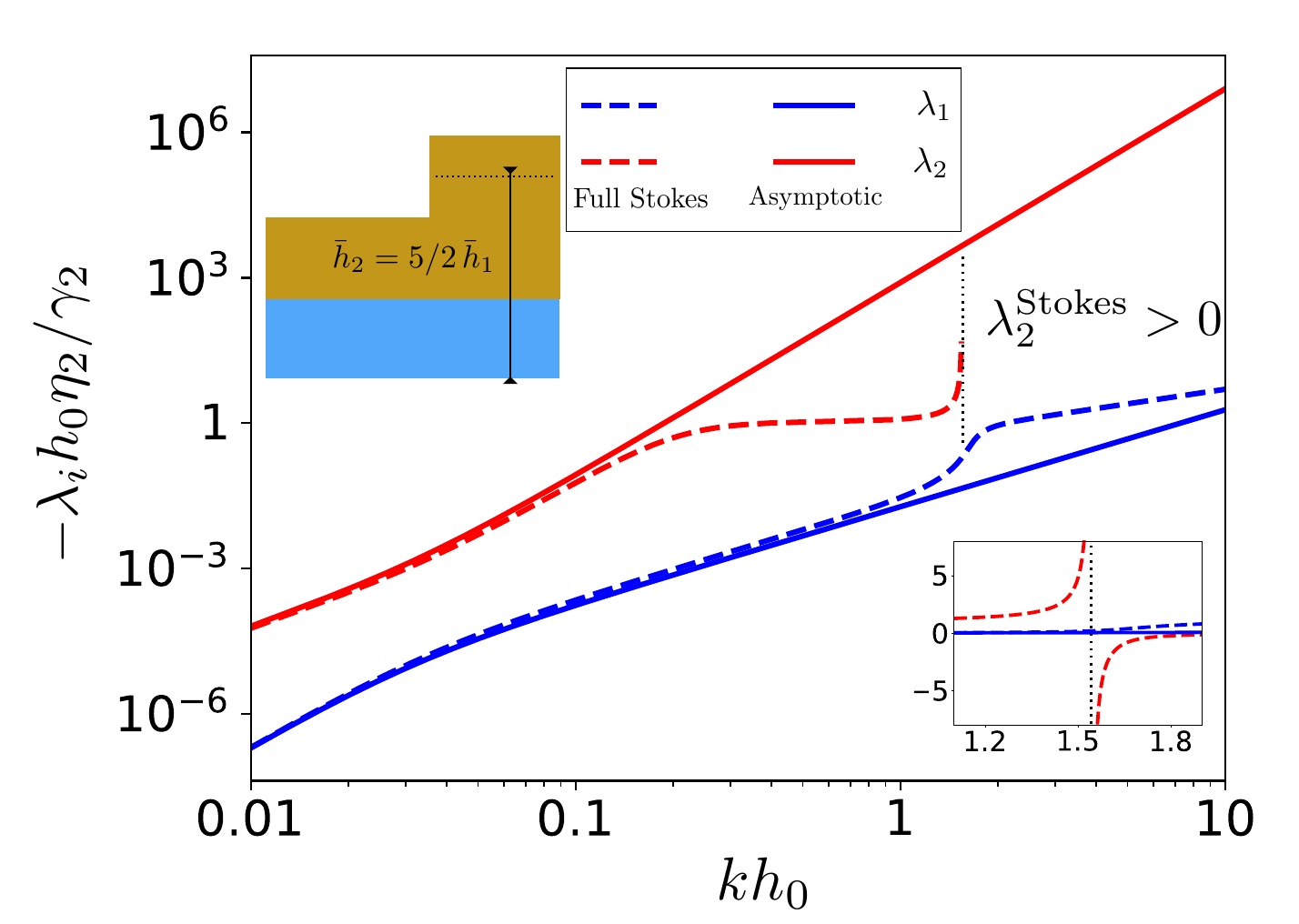}
\caption{Dimensionless decay rates for the \textit{full-Stokes model} ($\lambda_i^\textrm{Stokes}$) and \textit{asymptotic model} ($\lambda_i$) versus the dimensionless wave number $kh_0$, in the experimental configuration where $\bar{h}_2 = 5/2 \bar{h}_1 = 5/2 h_0$ (see schematic in the top left inset). The bottom right inset displays a zoom of the same curves near $k h_0 = 1.5$, and is plotted with linear scales.}
\label{fig:decayrate}
\end{figure}
Figure~\ref{fig:decayrate} displays the normalized decay rates as functions of the normalized wavenumber in both the \textit{full-Stokes model} and the \textit{asymptotic model}, with the same dimensionless parameters as in Fig~\ref{fig:profiles_topbot}, \textit{\textit{i.e.}} $\mu = 1.1\times 10^{-4}$ and $\Gamma = 0.053$. The two models agree in the small-wave-number limit, $kh_0 \rightarrow 0$. At an intermediate wave number $kh_0 \sim 1.544$, the determinant of the matrix $\mathbf{s}^\textrm{Stokes}$ changes sign and therefore one of the two eigenvalues, $\lambda_2^\textrm{Stokes}$, becomes positive at larger wave numbers (see bottom right inset of Fig.~\ref{fig:decayrate}). Thus, in the \textit{full-Stokes model}, the large wave numbers are unstable and grow with time: the interface perturbation diverges as time increases. This is not physical as capillarity is the only driving force and acts to stabilize the interface. We suspect that non-linear terms in the stress balances at interfaces -- neglected so far -- will regularize this behaviour. 

\subsection{Particular case: equal average layer thicknesses}
\begin{figure}[h]
\centering
\includegraphics[width = 0.95\textwidth]{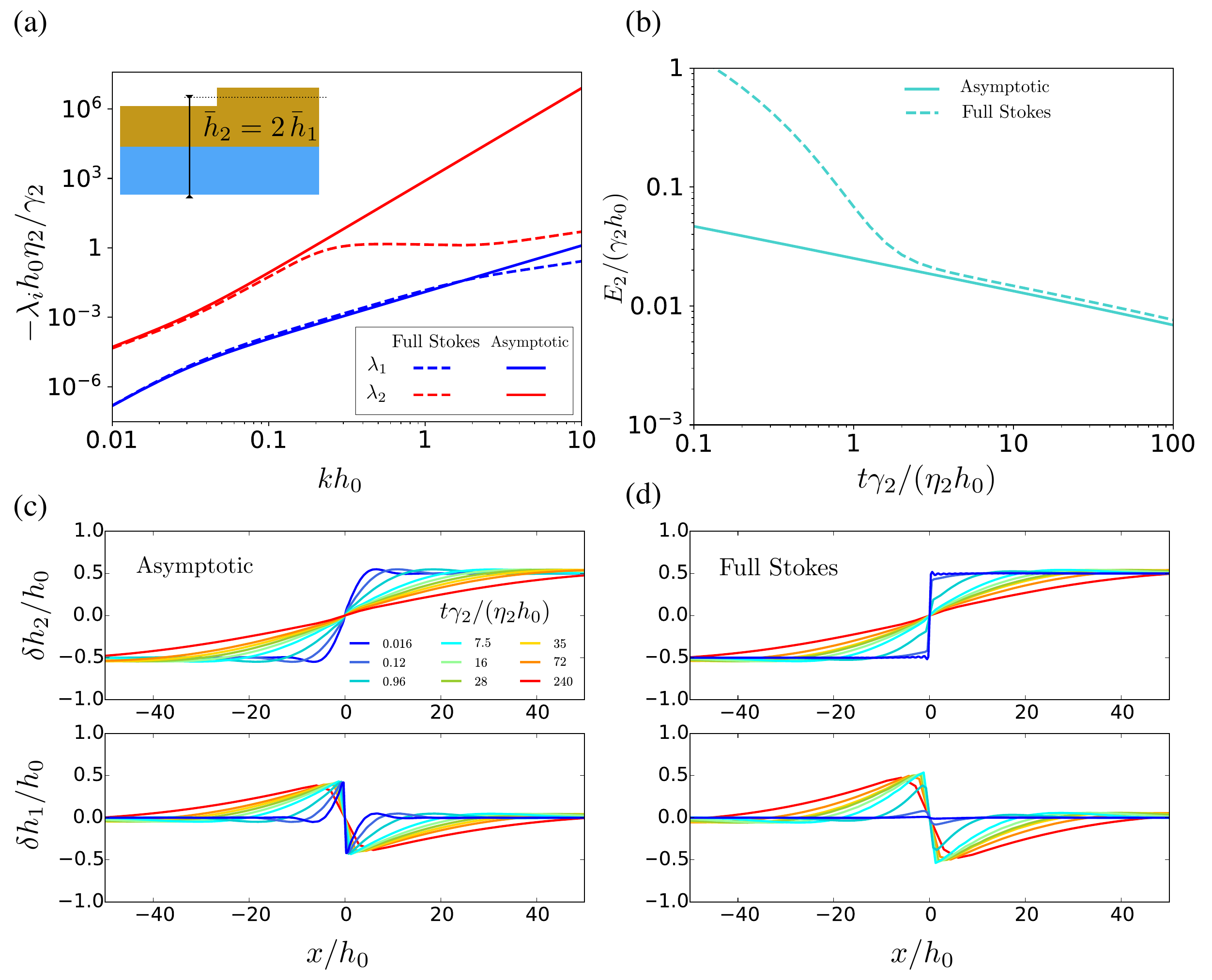}
\caption{Comparison of the \textit{full-Stokes model} and the \textit{asymptotic model} in the specific case where $\bar{h}_2=2\bar{h}_1$. (a) Dimensionless decay rates for the \textit{full-Stokes model} ($\lambda_i^\textrm{Stokes}$, dashed lines) and \textit{asymptotic model} ($\lambda_i$, solid lines) versus the dimensionless wave number $kh_0$. (b) Normalized excess capillary energies per unit length $E_2/(\gamma_2 h_0)$ of the liquid-air interface, as functions of dimensionless time, as evaluated from the small-slope expression $E_2 \simeq \gamma_2 \int \textrm{d}x \, \delta h_2'^2/2$, for both models and for the same parameters as in Fig.~\ref{fig:energy}(a) at 150$^{\circ}$C. (c) (resp. (d)) Interface perturbation profiles $\delta h_i$ in the \textit{asymptotic model} (resp. \textit{full-Stokes model}). The colors indicate the same experimental times as in Fig.~\ref{fig:profiles_topbot}.}
\label{fig:comparison_StokesAsymptotic}
\end{figure}
We found empirically that the instability described in the previous section is not present when the mean thicknesses of the two layers are equal, which amounts to $\bar{h}_2 = 2 \, \bar{h}_1$. Therefore, we can compare in a more complete manner the two models in this case. Figure~\ref{fig:comparison_StokesAsymptotic}(a) displays the normalized decay rates as functions of the normalized wavenumber in both the \textit{full-Stokes model} and the \textit{asymptotic model}. We no longer observe any positive decay rate in the \textit{full-Stokes model}. At small wave number, which means in the small-slope limit, we recover the previous statement which is that both models are consistent with each other. In Fig.~\ref{fig:comparison_StokesAsymptotic}(b), the normalized excess capillary energies per unit length of the liquid-air interface for both models are plotted as functions of dimensionless time. In the long-time limit, when the step has levelled sufficiently such that the typical slopes of the interfaces are much smaller than unity, we find an excellent agreement between both models. However, at short times, the profile slopes are close to unity and thus vertical flows and non-linear terms play a significant role. Therefore, at short times the \textit{full-Stokes model}, which accounts for vertical flows, differs from the \textit{asymptotic model}. We point out that the excess capillary energies per unit length $E_2$ of the liquid-air interface, from both models, are systematically computed with the small-slope expression $E_2 \simeq \gamma_2 \int \textrm{d}x \, \delta h_2'^2/2$ which is not necessarily valid at short times. The exact expression should be used instead to make direct comparisons with experiments at short times. The discrepancy between the two models at short times is illustrated on the interface perturbation profiles in Figs.~\ref{fig:comparison_StokesAsymptotic}(c) and (d), that would correspond to an experiment with the same material properties as in Fig.~\ref{fig:profiles_topbot} but with equal mean thicknesses. Interestingly, we observe similar short-term characteristics in the \textit{full-Stokes model} as the one observed experimentally: a small sharp feature near the step and the deformation growth of the liquid-liquid interface. 

\subsection{Case of a large viscosity ratio}
\label{Appendix_LargeViscosity}
\begin{figure}[H]
\centering
\includegraphics[width = 0.5\textwidth]{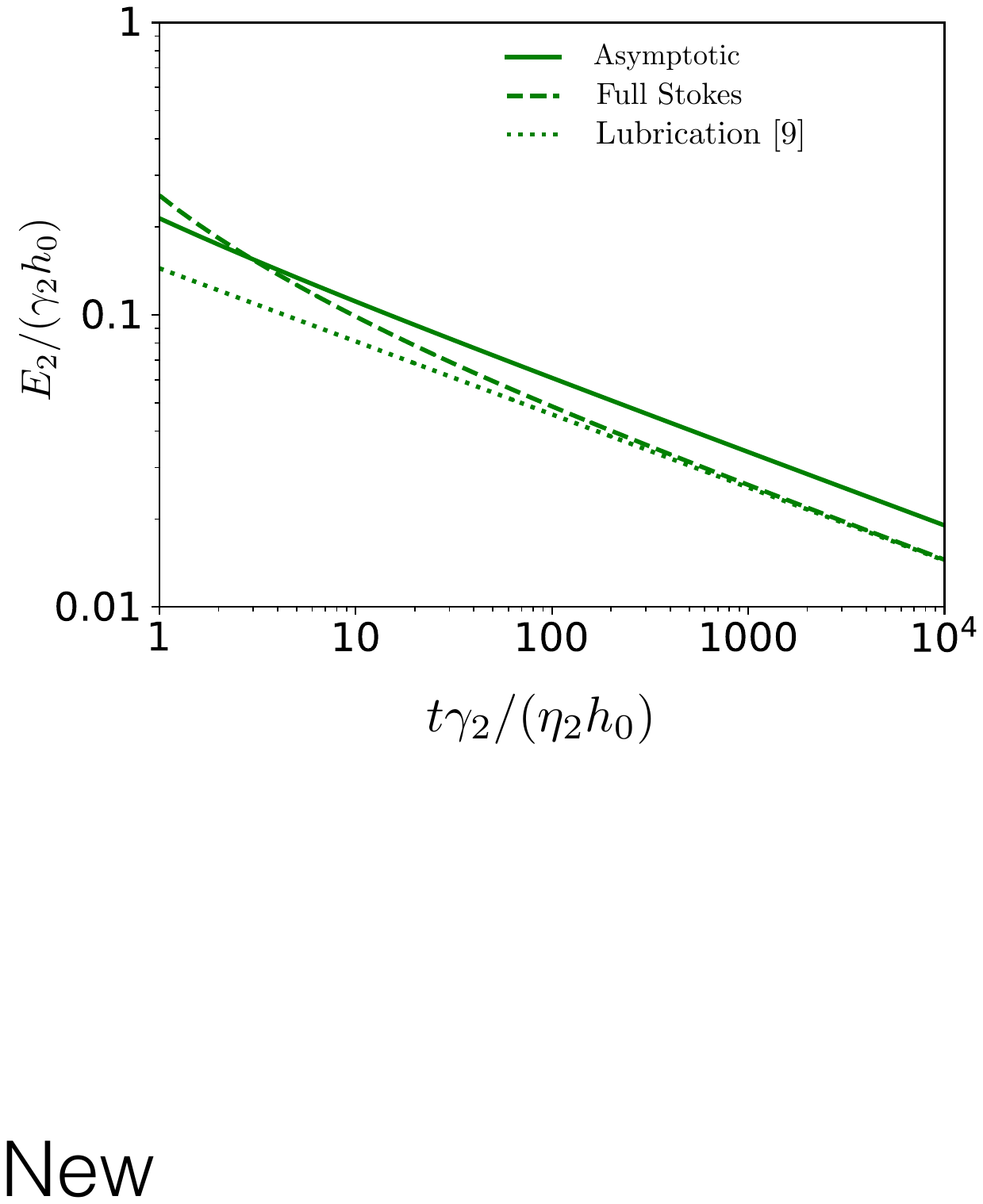}
\caption{Normalized excess capillary energies per unit length $E_2$ of the liquid-air interface as functions of dimensionless time, as evaluated from the small-slope expression $E_2 \simeq \gamma_2 \int \textrm{d}x \, \delta h_2'^2/2$, for the three models indicated, and for $\mu = 14$, $\gamma = 0.053$, with equal mean thicknesses $\bar{h}_1 = \bar{h}_2$.}
\label{fig:energy_StokesAsymptotic}
\end{figure}

We consider the $\mu\gg1$ case. In Fig.~\ref{fig:energy_StokesAsymptotic}, the normalized total excess capillary energies per unit length as functions of dimensionless time are plotted for the two models described above. We also add for comparison the perturbative solution of a two-layer \textit{lubrication model} (see \cite{jachalski2014} for a complete derivation in a more general case with weak slip):
\begin{subequations}
\begin{equation}
\frac{\partial h_1}{\partial t} = \left[-\gamma_2  \frac{h_1^2(3h_2 - h_1)}{6\eta_1}  h_2''' - \gamma_1 \frac{h_1^3}{3\eta_1} h_1''' \right]',
\end{equation}
\begin{equation}
\frac{\partial (h_2 - h_1)}{\partial t} =  \left\{ -\gamma_2 \left[ \frac{(h_2 - h_1)^3}{3\eta_2}   + \frac{h_1 (h_2- h_1)(h_2 - h_1/2)}{\eta_1}\right] h_2''' -\gamma_1  \frac{h_1^2 (h_2-h_1)}{2\eta_1} h_1''' \right\}'.
\end{equation}
\end{subequations}
We observe that the \textit{asymptotic model} is no longer in accordance with the \textit{full-Stokes model} in the large-time limit, while the \textit{lubrication model} is. Indeed, when the viscosity of the bottom layer is comparable to or larger than the one of the top layer, \textit{\textit{i.e.}} $\mu \gtrsim 1$, the \textit{asymptotic model} is no more valid as it neglects shear terms in the top layer with respect to elongational ones.

\section{Energy balance}
\label{AppendixEnergy}
In this section, we derive the energy balance in Eq.~\eqref{Equation:EnergyBalance} from the \textit{asymptotic model}. In the limit of small slopes, the excess capillary energies per unit length of the two interfaces are:
\begin{subequations}
\begin{equation}
E_2 = \frac{\gamma_2}{2} \int \textrm{dx}\,  h_2'(x)^2,
\end{equation}
\begin{equation}
E_1 = \frac{\gamma_1}{2} \int \textrm{dx} \,  h_1'(x)^2.
\end{equation}
\end{subequations}
We can derive these quantities with respect to time and get:
\begin{subequations}
\begin{equation}
\partial_t E_2 = \gamma_2 \int \textrm{dx}\,  h_2' \partial_t h_2' = - \gamma_2 \int \textrm{dx}\,  h_2'' \partial_t h_2,
\end{equation}
\begin{equation}
\partial_t E_1 = \gamma_1 \int \textrm{dx} \,  h_1' \partial_t h_1' = - \gamma_1 \int \textrm{dx} \,  h_1'' \partial_t h_1.
\end{equation}
\end{subequations}
The second equalities are obtained after integrating by parts. We can then use Eqs.~\eqref{GovEq1} and~\eqref{GovEq2}, which leads to:
\begin{subequations}
\begin{equation}
\partial_t E_2 = - \gamma_2 \int \textrm{dx}\,  h_2'' \left\{ \partial_t h_1 - [(h_2 - h_1) u_2]'  \right\} = - \gamma_2 \int \textrm{dx}\,  h_2'' \left\{ -\left( - p_1' \frac{h_1^3}{12\eta_1} +\frac{h_1 u_2}{2} \right)' - [(h_2 - h_1) u_2]'  \right\},
\end{equation}
\begin{equation}
\partial_t E_1 =  \gamma_1 \int \textrm{dx} \,  h_1'' \left( - p_1' \frac{h_1^3}{12\eta_1} +\frac{h_1 u_2}{2}  \right)'.
\end{equation}
\end{subequations}
We then integrate by parts:  
\begin{subequations}
\begin{equation}
\partial_t E_2 =  \gamma_2 \int \textrm{dx}\,  h_2''' \left\{ -\left( - p_1' \frac{h_1^3}{12\eta_1} +\frac{h_1 u_2}{2} \right) - [(h_2 - h_1) u_2]  \right\},
\end{equation}
\begin{equation}
\partial_t E_1 =  - \gamma_1 \int \textrm{dx} \,  h_1''' \left( - p_1' \frac{h_1^3}{12\eta_1} +\frac{h_1 u_2}{2}  \right).
\end{equation}
\end{subequations}
Introducing the total excess capillary energy per unit length $E = E_1 + E_2$, one gets:
\begin{equation}
\partial_t E =  - \int \textrm{dx}\,  \left[\gamma_2 h_2'''(h_2 - h_1)\right]  u_2  + \int \textrm{dx}\, p_1' \left( - p_1' \frac{h_1^3}{12\eta_1} +\frac{h_1 u_2}{2}  \right).
\end{equation}
We can then use Eq.~\eqref{GovEq3} to replace the term in square brackets:
\begin{equation}
\partial_t E =  - \int \textrm{dx}\,  \left\{ -\frac{p_1' h_1}{2}  - 4 \eta_2 [u_2' (h_2-h_1)]' + \eta_1 \frac{u_2}{h_1} \right\}  u_2  + \int \textrm{dx}\, p_1' \left( - p_1' \frac{h_1^3}{12\eta_1} +\frac{h_1 u_2}{2}  \right).
\end{equation}
This can be further simplified and after another integration by parts of the term in $[u_2' (h_2-h_1)]' \, u_2 $, one gets:
\begin{equation}
\partial_t E = - \int \textrm{dx}\, 4\eta_2 \, (h_2 - h_1)  u_2'^2 - \int \textrm{dx}\,  \eta_1 \frac{u_2^2}{h_1}  - \int \textrm{dx}\, \frac{ p_1'^2 \, h_1^3}{12\eta_1}.
\end{equation}

\bibliography{biblio}
\end{document}